# Control of Renewable Energy Communities using AI and Real-World Data

Tiago Fonseca[1,2*], Clarisse Sousa[1,2], Ricardo Venâncio[1,2], Pedro Pires[1,2], Ricardo Severino[1,2], Paulo Rodrigues[3], Pedro Paiva[4], Luis Lino Ferreira[1,2*]
INESC-TEC[1], Polytechnic of Porto - School of Engineering[2], ICharging[3], CleanWatts[4]
Porto, Portugal
{calof*, llf*}@isep.ipp.pt

*Abstract*— The electrification of transportation and the increased adoption of decentralized renewable energy generation have added complexity to managing Renewable Energy Communities (RECs). Integrating Electric Vehicle (EV) charging with building energy systems like heating, ventilation, air conditioning (HVAC), photovoltaic (PV) generation, and battery storage presents significant opportunities but also practical challenges. Reinforcement learning (RL), particularly Multi-Agent Deep Deterministic Policy Gradient (MADDPG) algorithms, have shown promising results in simulation, outperforming heuristic control strategies. However, translating these successes into real-world deployments faces substantial challenges, including incomplete and noisy data, integration of heterogeneous subsystems, synchronization issues, unpredictable occupant behavior, and missing critical EV state-of-charge (SoC) information. This paper introduces a framework designed explicitly to handle these complexities and bridge the simulation-to-reality gap. The framework incorporates EnergAIze, a MADDPG-based multi-agent control strategy, and specifically addresses challenges related to real-world data collection, system integration, and user behavior modeling. Preliminary results collected from a real-world operational REC with four residential buildings demonstrate the practical feasibility of our approach, achieving an average 9% reduction in daily peak demand and a 5% decrease in energy costs through optimized load scheduling and EV charging behaviors. These outcomes underscore the framework's effectiveness, advancing the practical deployment of intelligent energy management solutions in RECs.

## I. INTRODUCTION

Modern smart buildings and energy communities are increasingly integrating distributed energy resources (DERs) such as solar photovoltaics (PV), battery storage, and electric vehicle (EV) charging infrastructure. Collectively, buildings account for approximately 32% of global energy consumption and 34% of energy-related $CO_2$ emissions, underscoring their pivotal role in climate mitigation efforts [1]. EV adoption continues to accelerate; in 2024, global sales of electric vehicles, including plug-in hybrids, surged by 25[2].

This growth presents challenges, such as increased strain on local electrical grids, particularly in areas with high EV concentrations, potentially necessitating costly upgrades to transformers and distribution networks. Many existing buildings, lack the infrastructure to support efficient EV charging, leading to energy inefficiencies and further stress building power supplies [3].

However, integrating EV charging with building heating, ventilation, and air conditioning (HVAC) systems and on-site renewables within Renewable Energy Communities (RECs) presents opportunities to mitigate these challenges by reducing peak demand, lowering overall energy costs, and providing valuable grid services [4], [5]. Nonetheless, coordinating these heterogeneous subsystems in real-time remains complex, owing to their diverse operational dynamics, system constraints, and the substantial influence of human behavior, such as occupants' comfort preferences and unpredictable EV usage patterns.

Reinforcement learning (RL) has emerged as a promising approach for such sequential decision-making problems in building energy management [6]. By learning control policies from interaction with the environment, RL can in principle optimize multi-objective outcomes (cost, comfort, etc.) better than rule-based or schedule-driven strategies. In particular, multi-agent deep RL algorithms allow decentralized control of multiple interactive assets (such as flexible loads in different buildings) with a coordinated global objective [7]. In the CityLearn simulator [8] benchmark environment, a digital twin for multi-building energy control, RL agents have outperformed baseline heuristics in coordinating batteries and flexible loads to achieve demand response objectives [9]. These successes highlight the potential of RL to unlock demand flexibility in smart buildings and energy communities.

Yet, translating these simulation successes to real-world deployments is non-trivial, as most of the implementations remain using controlled, treated and pre-processed datasets, which removes any real-world variances and difficulties, that real-world deployments need to handle [10]. Field trials of advanced control algorithms reveal many practical challenges, that can degrade an RL (or other control methods) performance [10], [11]. This Simulation-to-Reality Gap and challenges are an object of study of this paper and are detailed in Section II. Moreover, most of the research only focus on the algorithm, and not on the overall framework making it possible (collecting data, treating it in real-time, connecting to APIs and finding solutions to difficult to get data (most of the times simulated) [11].

As such, this paper builds on our previous research involving *EnergAIze* [12], a Multi-Agent Deep Deterministic Policy Gradient (MADDPG) reinforcement learning algorithm for managing RECs. Previous work proven that the *EnergAIze* was effective and surpassed comparative studies in the state-of-the-art when applied to a pre-prepared simulation dataset. In this paper we describe the work we have been developing to connect and surround *EnergAIze* with a framework specifically designed to handle the complexities associated with real-world REC operations and data collection, managing and mitigating challenges related to incomplete data, synchronization errors, and missing EV-related data (identified in Section II), and bringing the solution closer to a possible real-world deployable solution. The framework is used to collect real-world data of an operational REC composed of four residential buildings, equipped with photovoltaic (PV) systems, battery storage, and EV chargers. Results, still in simulation, but made with real-world data are shown to prove the framework practical feasibility and effectiveness, and its readiness to get to the next step of deployment.

This paper contributes to bridging the existing gap between theoretical research and practical deployment by proposing an integrated framework that presents some solutions to the often-underreported engineering hurdles of real-world deployment. The key contributions can be summed up as:

- a detailed discussion of the Simulation-to-reality Gap and challenges, with perspectives from both the state of the art and from our practical experience;
- integrating *EnergAIze* multi-agent MADDPG (Multi-Agent Deep Deterministic Policy Gradient) control for co-optimizing building HVAC, battery, and EV charging into a more real-world ready framework that handles data collection, processing, API connection, among others;
- preliminary results from using real-world data collected in the framework.

The remainder of the paper is organized as follows. Section 2 reviews the Simulation-to-Reality Gap and challenges associated with real-world deployment of REC management solutions. Section 3 details the proposed framework, including the multi-agent MADDPG controller and system architecture. Section 4 presents the results and discussion from the four real-world buildings. Finally, Section 5 concludes with key lessons and future work directions for scaling and improving RL in building energy management.

## II. SIMULATION-TO-REALITY GAP AND CHALLENGES

Despite extensive simulation research, there have been only limited deployments of RL controllers in real buildings to date. A few pioneering field implementations have started to emerge. Kurte et al. (2020) trained a deep Q-network (DQN) agent in simulation to control HVAC in a test house, then deployed the policy in a real single-family home [13]. Authors reported that the pre-trained RL policy could be applied in the real building and maintained comfort while reducing energy use, though the training had to be done offline due to data limitations. Naug et al. (2020) went further by implementing a continual learning RL framework in a campus building: they periodically updated the RL policy on new data to adapt to non-stationary operating conditions[14]. This approach, which they called a "relearning framework," was successfully tested in a three-story university building's climate control system, showing the feasibility of online adaptation. Another example is the work by Luo et al. (2022), who deployed deep RL agents to control cooling systems in two large commercial buildings, achieving reported energy savings of 9–13% in HVAC electricity consumption [15]. Their multi-stakeholder project required addressing numerous practical issues, and the authors summarized open challenges including how to evaluate control policies in situ, how to learn from limited real data, dealing with multi-timescale dynamics, enforcing safety constraints, and handling varying operating scenarios. These real-world trials confirm that RL can yield energy efficiency gains in practice, but also illustrate the engineering complexity of implementation, and that none of the approaches were generic enough to be deployed at the level of dozens of family homes, each with different assets and complexities. Moreover, to the best of the authors knowledge, specifically no MARL algorithm has been applied into a real world setting. Authors in [10] reported the challenges and gaps for MARL algorithms, but not for their real-world deployment.

Park et al. (2021) critically reviewed field implementations of occupant-centric controls and found that current building information technology infrastructure is often not supportive of advanced algorithms like RL in terms of sensor coverage, data accessibility, and controllability of equipment [16]. Researchers have resorted to creative workarounds, such as installing custom IoT devices (Arduino/Raspberry Pi) to interface with legacy equipment. Chan et al. (2023) discuss similar integration hurdles, noting the lack of standardized protocols and prevalence of vendor-specific systems in building management which complicate IoT deployments [17].

Here we highlight some of the challenges noted by literature and complete with our own input of working within this area. Key issues include:

- **Noisy and Incomplete Data:** Operating a REC in reality means dealing with imperfect data streams from IoT sensors, smart meters, inverters, etc. Missing readings, communication dropouts, and asynchronous timestamps are common; consequently, the agents face partial observability and must act on stale or noisy inputs. Most RL research to date assumes well-synchronized and pre-processed data, so algorithms often lack robustness to the data quality issues encountered in the field. A controller must therefore be resilient to sensor faults and estimation errors if it is to maintain reliable performance outside the lab. The framework presented in this paper, has a robust data pipeline,the *Percepta* component (Section III.b), that continuously clean, validates, and imputes sensor data in real-time so that the controller is not derailed by anomalies missing value.

- **Unmodeled Disturbances:** Simulations rarely capture the full variability of human behavior and other stochastic disturbances. In a real REC, occupants may change their consumption habits, thermostats, or EV usage unpredictably, and weather events or equipment faults can introduce dynamics never seen during training. Such non-stationarities violate the simulator's assumptions and can lead to severe performance degradation when an RL policy is deployed. For instance, an agent trained on a fixed EV charging schedule may falter if an EV is suddenly driven at an unexpected time [18]. Literature acknowledges that RL controllers struggle with transfer to altered environments (e.g. a building retrofit or different occupant behavior) without additional adaptation [10]. Hence, training with real world data, without any processing is a must for real world deployment robustness. At the presented framework, data for training is passed as it is collected, with the variance and disturbances from the real world, as for example an EV owner states that is leaving at 8am but effectively leaving at another time.

- **Heterogeneous System Integration:** RECs comprise diverse assets and proprietary subsystems, solar PV inverters, battery management systems (BMS), EV chargers (often with vendor or manufacturer-specific protocols), HVAC and water heating controls, etc. Integrating an RL agent with all these heterogeneous devices in a real building or microgrid is non-trivial. Many commercial energy management systems are "closed" and do not offer open APIs for real-time control, forcing developers to design custom interface middleware or reengineering proprietary ones. Prior works have noted that current building automation technology is not yet conducive to advanced controllers like RL, due to fragmented sensor networks and limited interoperability. This lack of a unified integration layer means that deploying a generic enough MARL solution requires significant engineering effort to interface with different data formats, communication buses, and vendor hardware. The *Percepta* component (Section III.b), developed to serve as a bridge between the data sources and *EnergAIze*, and the *PulseCharge* module (Section III.d), which standardizes the Cars API to obtain information (e.g., state of charge), work together for the unification of data as well as to get data from EVs and proprietary systems.

- **Real-Time Coordination and Latency:** In live control settings, decisions often need to be made and executed within seconds or milliseconds, especially for fast-responsive resources or to track grid signals. If the MARL controller relies on cloud computing or long communication loops, latency can hinder its ability to provide timely actions. Network delays or outages pose a risk when control is not local. A practical system should minimize dependency on always-connected cloud services and instead leverage edge computing to keep response times low and ensure continuity during network disruptions. In the framework presented in this paper, deployment on the edge (local controllers) is favored over cloud-only implementations, to reduce latency and dependency on internet connectivity. By running the MARL optimization on a local embedded device or gateway at each site (as *EnergAIze* suggested with per-dwelling nodes), the system can react quickly to changes and continue operating during network disruptions, all while keeping sensitive data on-site for privacy. The *Percepta* component also supports decentralized execution (Section III.b).

- **Fail Safe Mechanisms:** In real-world energy systems, the consequences of faulty control actions can be serious ranging from equipment damage to user discomfort or even safety hazards. Unlike simulations, where mistakes have no physical repercussions, deployed RL controllers must always ensure safe operation. However, most RL algorithms are not inherently safe by design and may explore unsafe actions if not properly constrained (even if they exhibited good results on training). This becomes particularly critical in decentralized or MARL settings, where coordination failures between agents can compound risks. Literature has pointed out the need for mechanisms that detect when the policy deviates from safe behavior or when external conditions change beyond the agent's learned experience [19]. In practical implementations, fallback strategies are essential, such as defaulting to rule-based control, safe standby modes, or operator-defined setpoints during system anomalies, unexpected sensor values, or prolonged communication failures. The proposed framework includes a dedicated Fail-Safe Supervisor component, which monitors agent actions and intervenes when pre-defined safety thresholds are breached. For example, if the an EV is not charging as expected, the supervisor can override RL outputs and apply a conservative backup control from an RBC. Additionally, all decisions are logged and auditable, enabling post-event evaluation and ensuring transparency and future analysis.

- **User Acceptance and Comfort Preservation:** Ultimately, any automated energy control system in a community must be acceptable to its human users. In theory, RL can learn nuanced strategies, but if those strategies inconvenience residents or require constant user input, the deployment will fail [16]. User-transparent operation is crucial, the system should manage energy flexibly in the background without frequently asking occupants for manual decisions or data (e.g. prompting for EV battery state of charge or demand flexibility on a daily basis). Prior pilot studies that attempted to solicit occupant feedback (for example, via a smartphone app to adjust comfort preferences) found very low engagement, undermining the effectiveness of the RL approach [20]. Likewise, maintaining comfort is non-negotiable: aggressive control actions that violate thermal or appliance usage comfort will lead occupants to override or abandon the system. Thus, modern RL controllers must explicitly respect hard constraints like temperature and device availability limits and balance objectives such that energy goals are achieved without noticeable sacrifice to comfort. Lack of consideration for these human factors in many RL prototypes has been a major barrier to real-world adoption [20]. The presented framework MARL controller works autonomously in the background, and any user inputs (like comfort settings or vehicle needs) are either gathered implicitly or only required to be input by the user at infrequent intervals. Advanced forecasting and learning allow the system to anticipate user needs (e.g. likely EV departure times or preferred room temperatures) so that it does not routinely ask users for information. In the presented framework, the *PulseCharge* component (Section III.d) and the flexibility prediction component (Section III.e) are put in place to reduce the user needed interaction.

Most existing reinforcement learning research for smart buildings and energy communities falls short of addressing the above challenges. Studies typically report results on idealized datasets or simulations, for example, using historical data with all values neatly filled in, or assuming perfectly rational occupant models, which fails to reflect the messy conditions of real deployments. As a result, controllers that excel in academic benchmarks often struggle when confronted with noisy sensors, unforeseen events, or hurdles integration outside the laboratory. This indicates that focusing solely on algorithmic performance in simulation is insufficient, the system-level context must be accounted for. To bridge this simulation-to-reality gap, our proposed approach argues for a complete system integration framework surrounding the *EnergAIze* algorithm. In such a framework, the learning agent is just one component of a broader solution that also encompasses reliable data acquisition, real-time processing, and seamless user interaction.

### III. Architecture

The *EnergAIze* framework (inheriting its name from the algorithm at its core) has been developed to enable intelligent, energy flexibility management within RECs. It achieves this by equipping each building, either residential, commercial, or any other type, with cyber-physical components. Figure 1 illustrates a high-level overview of the framework's architecture, showcasing the interaction between its core components and REC physical assets.

Each building in a real-world REC is unique in its composition and may include a variety of DERs, such as PV panels, battery energy storage systems (BESS), EV chargers, heat pumps, and other flexible loads (e.g., HVAC systems, electric water heaters, washing machines). These assets can differ significantly in terms of their manufacturer, communication protocols, and API accessibility. As depicted in Figure 1, control and monitoring are achieved through multiple proprietary and standardized interfaces. For instance, EV charging infrastructure may be managed via the cloud services of the manufacturer's platform (e.g., my.i-charging), while other assets are integrated through third-party platforms such as the Cleanwatts (CW) Cloud Platform, a company partner in this work. Additional IoT devices may operate independently or connect via their own dedicated APIs. Ultimately, the degree of integration and control depends on each asset's IoT capabilities, with partial observability or fallback mechanisms employed when direct control is not feasible.

To support real-world, privacy-preserving decision-making, we design the framework so that each building hosts an instance of the needed components at the edge (e.g., running on an edge-fog device at each prosumer house). Each one of the edge nodes (highlighted yellow in Figure 1) integrate a local control agent, data interfaces, safety mechanisms, and forecasting tools. These instances operate in a decentralized manner, meaning each household independently optimizes its own consumption and flexibility while also contributing to REC-wide goals such as peak shaving, energy sharing, or improved renewable utilization. The architecture is built to be modular and interoperable, allowing integration with heterogeneous systems and easy extension to new asset types.

The chosen design fulfills the objectives and tackles the identified challenges (Section II) in several ways, as it is decentralized (computing at the edge, not a monolithic controller, improving resilience to system wide downtimes), it is interoperable (through standardized data interfaces with building/vehicle systems), and it operates in soft real-time (agents compute actions quickly as data arrives, and actions are applied to devices). Additionally, by keeping computation local and data sharing minimally, the system inherently supports data privacy and has a smaller cyber-attack surface (only aggregated or necessary data goes on the network, and each agent can firewall its local device controls). This architecture is extensible to larger communities, as one can add more agent nodes for new prosumers, and the *Percepta* can scale horizontally to handle more data streams, since each agent largely functions independently aside from light communication with the cloud to obtain energy prices.

At the core of each building's framework lies the *EnergAIze* Algorithm (shown in green in Figure 1) and the *Percepta* module (orange). The *Percepta* acts as the local decision-making hub, interfacing with both IoT-enabled devices and associated cloud platforms. It collects and processes heterogeneous observational data (blue arrows), including real-time sensor readings and predictions from specialized forecasting and data collection components: the *Consumption and Production Prediction*, *Flexibility Prediction*, and *Pulse Charge* components.

This information is relayed to the *EnergAIze* Algorithm, which computes optimal control actions. These actions are then routed back through the *Percepta*, where a *Fail-Safe Supervisor* module verifies their feasibility and safety before dispatching them to the appropriate actuators via the available communication interfaces. All interactions and resulting actions are logged in a local database to support monitoring, accountability, and post-hoc analysis. In each of the sub-sections below, we summarize the key modules of the framework.

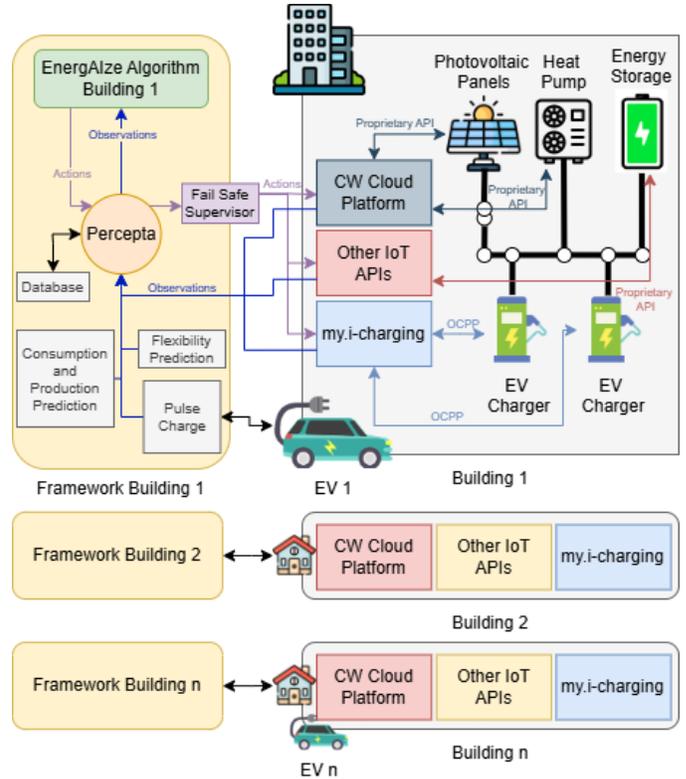

*Figure 1 - Framework Architecture*

*A. EnergAIze Agent*

At the heart of each household's framework is a local MARL agent, implemented the algorithm presented in [12]. It receives input from the home's *Percepta* (consumption, PV production, battery SoC, EV status, etc.) and outputs control decisions, such as when to charge an EV, shift HVAC setpoints, or export battery energy. These decisions are made on a fixed schedule (e.g., every 15 minutes) and are tailored to the habits and assets of the specific prosumer/household. The agents are trained in a centralized simulator but deployed locally, enabling decentralized execution and preserving privacy. For more details on the algorithm and how it compares to other state-of-the-art approaches, please refer to our previous work [12].

*B. Percepta*

This component is the framework's communication, data normalization, data synchronization, data cleaning and data preparation orchestrator. It operates as a system that, at regular intervals (defined by the agent's decision-making frequency, e.g.: 15 minutes), requests or receives data, establishes meaningful correlations (i.e.: despite data coming from different providers, it makes sense of these and gathers it all under one object) and generates an encoded message in the format required by the inference algorithm to enable decision-making. Through this process, *Percepta* addresses several of the challenges identified in previous sections.

The types of data handled by this component include: (i) sensor and meter data from buildings (such as total building load, PV inverter output, battery state-of-charge from the energy management system, etc.), (ii) environmental data like weather conditions and forecasts, and (iii) market signals such as electricity tariffs or grid requests (if applicable).

*Percepta* (depicted in Figure 2) is built with adapters for standard protocols, for instance, it can pull measurements from a building management system via BACnet or Modbus, retrieve charger status via OCPP (Open Charge Point Protocol), and call external web APIs for weather forecasts, for example the Cleanwatts Cloud Platform (illustrated at Step 1 in Figure 2). By using standard interfaces, the module ensures the system can integrate with existing infrastructure without custom rewiring, thus meeting the interoperability goal.

Then, *Percepta* also performs data preprocessing and quality control (Step 2): it timestamps and synchronizes data from different sources, filters out anomalies (e.g., sensor spikes), and fills in missing data points.

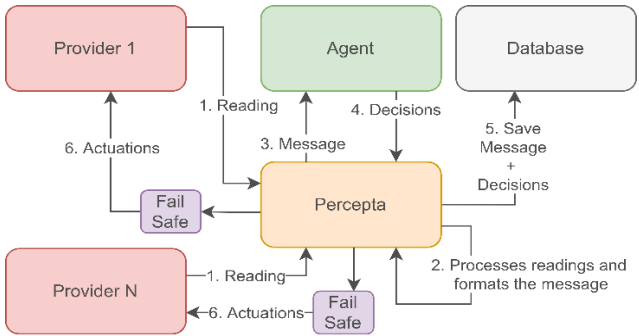

*Figure 2 - Percepta Flows*

For critical signals, if data is temporarily unavailable (network glitch or sensor fault), the *Percepta* can substitute it with a recent value, or a prediction based on historical patterns so that the agents can still make informed decisions. This resilience in the data layer addresses the issue of partial observability and noisy data.

After aggregating, formatting, and sending the data for inference, *Percepta* sends the observations into the Algorithm Agent (Step 3), and when a decision is taken (Step 4), it routes the resulting decisions from the algorithm to the appropriate endpoints (Step 6), ensuring that control actions are correctly applied to the respective end devices. In parallel, it stores all messages sent to the agent, along with the resulting decisions, in a database to enable future accountability, analysis and support continuous model retraining (Step 5).

*Percepta* runs at an edge/fog computer. All communication is encrypted (using TLS) to prevent eavesdropping or tampering, reflecting a cybersecurity-by-design approach.

*C. Fail Safe Supervisor*

In real-world deployments of RL systems for energy management, ensuring the safety, reliability, and robustness of control actions is of high importance as we have discussed at Section II. Even though *EnergAIze* agents are trained per default with safety constrains and Physics constrained RL, demonstrating strong performance in simulation, it may behave unpredictably when confronted with distributional shifts, sensor faults, or previously unseen scenarios in the physical environment, especially during the research and development phase we are in. These limitations necessitate the integration of explicit fail-safe mechanisms into the operational control loop to act as a safeguard against unsafe behavior.

To address this challenge, the proposed *EnergAIze framework* includes a dedicated *Fail-Safe Supervisor* (FSS) module as a core component of its architecture and integrated within *Percepta*. The FSS acts as a real-time gatekeeper that verifies actions proposed by the RL agent against a set of pre-defined safety constraints and fallback criteria. This component is responsible for monitoring, validating, and, if necessary, overriding the agent's decisions to ensure that all actions conform to safety, comfort, and operability boundaries.

For instance, if the agent attempts to charge an electric vehicle at a rate exceeding the hardware's rated capacity, or if it proposes discharging a home battery below a critical state-of-charge threshold, the FSS intervenes to prevent execution. In such events, the supervisor substitutes the unsafe action with a conservative, predefined fallback strategy, typically derived from rule-based control (RBC) logic or operator-defined setpoints. These fallback policies prioritize system stability and user comfort over optimization, ensuring graceful degradation of functionality during anomalies rather than complete failure.

*D. PulseCharge*

Accessing data from EVs, particularly across a vast set of manufacturers, presents several challenges. Each manufacturer typically implements proprietary communication protocols, data formats, and security mechanisms for interfacing vehicle subsystems such as the battery management system, telematics unit, and vehicle control modules. This lack of standardization complicates the integration process for third-party platforms aiming to retrieve vehicle data, such as *EnergAIze* framework or others that need data from the EV to another objective.

In September 2025, the European Commission's Data Act [21] is set to come into effect, resulting in transformative changes in data accessibility across the European Union. According to X, this regulation will impose new obligations on vehicle manufacturers, particularly for electric vehicles (EVs), by mandating the provision of open APIs and public documentation. As a result, the process of retrieving data from EVs will be streamlined, granting users and third-party service providers easier access to critical vehicle-generated information.

*Pulse Charge* motivation comes from the above challenges to provide *EnergAIze framework* with EV data before the EU Data Act comes into effect. It specializes in EV integration and gathers data such as SoC, charging power, and expected departure time, and delivers this information to the agent, via the *Percepta*. It enables the agent to treat the EV as a flexible storage asset and optimize its charging accordingly, respecting user's needs while taking advantage of renewable energy availability or grid signals. Since this is sensible data to the prosumer, ensuring secure and authorized access requires adherence to robust authentication protocols, such as OAuth 2.0, and compliance with data protection regulations.

Figure 1 illustrates how *PulseCharge* interfaces with both the EV (through the car's network or charger) and the *EnergAIze* agent. The user performs a one-time setup by accessing *PulseCharge* and creating an account, providing relevant information such as the identification of their residence. This step is crucial for linking the EV to the correct household. The user then selects the EV manufacturer, associating the vehicle with the *PulseCharge* platform. Since accessing EV data requires authorization, *PulseCharge* redirects the user to an external Authorization Server. There, the user logs in (their manufacturer application) and grants the necessary permissions for *PulseCharge* to access their EV information. Once consent is given, the Authorization Server redirects the user back to *PulseCharge* along with an authorization code. *PulseCharge* then exchanges this code for an access token, which it uses to securely request data from the user's EV. The EV responds by sending the requested information back to *PulseCharge*, allowing the user to interact with and view their vehicle data through the platform. The information is subsequently disseminated to the designated *Percepta* periodically (e.g., every minute).

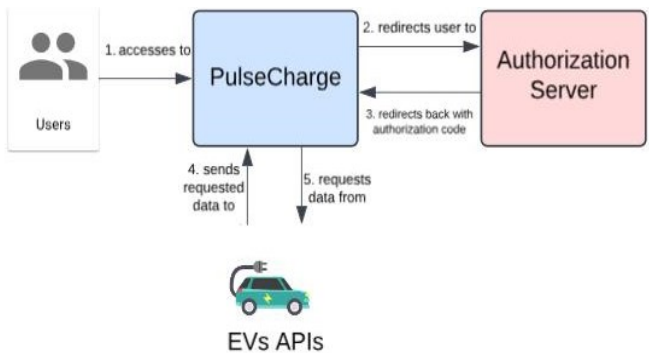

*Figure 3 – Pulse Charge Configuration and Operation*

### E. Prediction and Forecast Components

The *EnergAIze* framework includes two key forecasting modules: the *Consumption and Production Prediction* component and the *Flexibility Prediction* module. Together, they support proactive and data-informed decision-making by providing estimations of expected energy demand, production, and flexibility available at each building. These components play a role in enhancing the robustness of control actions and ensuring system resilience, particularly under conditions of missing or delayed sensor data.

The *Consumption and Production Prediction* module estimates short-term energy consumption and renewable energy production based on historical patterns, contextual data (e.g., weather forecasts), and behavioral routines. This information is instrumental for the control algorithm to anticipate demand peaks, optimize charging schedules, and balance the use of on-site resources. Additionally, in the event of sensor failures or connectivity issues, these predictions serve as a fallback mechanism to fill in data gaps, thereby maintaining system functionality and continuity, acting to solve one of the identified challenges in Section II.

The Flexibility Prediction component is responsible for estimating user specific data regarding their EVs charging patterns and flexibility. Its motivation comes from one fundamental aspect of successful RECs: the collaboration from the prosumers so the system may work according to their specific needs. However, too frequent real time requests on specific information, such as arrival time, initial or desired SoC, can become too intrusive, boring, causing overall user disengagement and lack of their participation in the REC.

This module originated from the need to ensure that each user plays their part in the REC, while still respecting the user's "personal space" and simplicity. This component is layered around the user's willingness to participate actively, in the absence of which it opts for a more passive approach, based on statistical models built with user particular historical data.

1. Flexibility Estimation (FE) – the component is pessimistic, i.e. at the beginning of each day it will assume that the user will not manually provide their preferences for day, therefore estimating the most likely set of values for the flexibility based on historical data.

2. User provided preferences – the previous estimation is overwritten by the active collaboration of the user, given its higher priority.

The FE functionality is to be modeled with caution. There are many variants that come into play when estimating the flexibility parameters of a particular user, such as the changing User Schedule, where flexibility parameters, such as time of arrival or SoC, are not the same for each day of the week, the Holidays and Vacations, requires its own FE, given how the user patterns can vary on these days and the Seasonal Changes, as depending on the time of the year, the user schedule may change, and the FE needs to adapt to the new patterns.

FE requires some additional attention to the Sample Size, as the bigger the sample and historical data is, the more reliable the model will be. In operation, the *FE* component receives a time stamp as input, representing the departure time of a user's EV, and build intervals, with a chosen level of confidence, for the most likely values of arrival time, charging duration and energy charged that provides to the *Percepta*.

A future resolution may discontinue this "trial and error" approach, with the use of a GPS system on each EV that is then used to connect to the framework and provide information about not only the time of departure, but also the SoC at departure, time of arrival and Soc at arrival. This ends up being non-intrusive, more precise and reduces computing resources usage.

## IV. RESULTS

Although the system has not yet been deployed in a live setting, all data used in this study originates from a real REC composed of four residential households, that have EVs, stationary batteries BESS, and PVs. Each of the four buildings features a 7.4 kWh EV charger, three include photovoltaic (PV) systems of varying sizes, and one house includes a 9.6 kWh battery energy storage system (BESS).

Over the course of 1.5 years, data was continuously collected from smart meters, EV chargers, and photovoltaic inverters via the REC company cloud and then collected to build the dataset through the available API. This dataset includes true operational characteristics, such as communication dropouts, asynchronous timestamps, and real Iberian market (OMIE) electricity pricing, providing a more realistic testing ground than synthetic datasets typically used in simulation research. A Simulation Scenario (SS) was constructed using this real-world data and converted into a format compatible with the CityLearn environment, extended with the EVLearn module [22] to support electric vehicle simulations. The simulator, filled with real-world data functioned mostly as a digital-twin. In the simulation, each dwelling pursued an individual cost-minimization objective, while the REC targeted peak demand reduction. KPIs were computed for the final deterministic episode (i.e., post-training execution) using normalized values against a baseline case with no intelligent energy management.

The following KPIs based on literature-supported objectives for energy flexibility, were used: Electricity Consumption (D), Measures imported energy from the grid; Electricity Price (C), Captures total energy cost, including price fluctuations; Zero Net Energy / Self-Consumption (Z) – Indicates the extent of energy self-sufficiency; Average Daily Peak (P), Measures average daily maximum energy use at REC level; Ramping (R), Captures fluctuations in energy consumption hour-to-hour.

*Table 1 - Results using Real-World Data*

| KPIs | B1 | B2 | B3 | B4 | REC |
|------|------|------|------|------|------|
| D | -3.14% | -1.05% | -0.96% | -2.10% | - |
| C | -7.24% | -2.23% | -4.16% | -5.81% | -4.86% |
| Z | +6.45% | +3.54% | +3.37% | +4.07% | +4.36% |
| P | - | - | - | - | -9% |
| R | - | - | - | - | -17% |

Table 1 presents normalized KPI results, comparing *EnergAIze's* performance after 15 training episodes to the baseline. All values are expressed as percentage improvement over the baseline scenario.

At the dwelling level, the framework successfully reduced grid-imported electricity, leading to greater intra-community energy sharing. This enabled cost reductions, especially in buildings B1 and B4, which also saw the most flexible usage of their BESS and EV charging schedules.

Notably, these cost savings were achieved without sacrificing user comfort or requiring manual input. Improvements in the Zero-Net Energy KPI across all buildings reflect an increase in local self-sufficiency. This effect stems from load shifting to align with higher PV generation and mutual energy exchanges among buildings within the REC.

At the community level, the framework controller significantly reduced peak demand by 9% and ramping fluctuations by 17%, both of which are crucial for grid stability and demand response participation. These results demonstrate that the controller not only optimizes individual behaviors but also achieves coordinated community-wide benefits.

## V. CONCLUSIONS

This paper presented a framework designed to bridge the simulation-to-reality gap in the deployment of intelligent energy management systems for Renewable Energy Communities (RECs). By integrating the *EnergAIze MARL controller* with robust data pipelines of the *Percepta* component, real-time EV data acquisition (via *PulseCharge*), the *flexibility prediction* component, the decentralized design, the *fail-safe* component and interoperable communication with third-party platforms (e.g., Cleanwatts and my.i-charging), the framework addresses many of the practical challenges that typically hinder real-world RL deployment, including noisy data, asset heterogeneity, and latency constraints.

Evaluation results based on 1.5 years of real-world data collected from an operational REC with four residential households showed encouraging performance. *EnergAIze* achieved a 9% reduction in peak demand, a 17% improvement in ramping smoothness, and average energy cost reductions of nearly 5%, all while maintaining self-consumption and user comfort. These results support the pursue of future work to transition from simulation using real world data to actual real-world deployment. The next phase involves live testing of the system in operational environments to validate control performance under real-time constraints and user behavior variability. This will allow us to assess robustness, user acceptance, and system responsiveness under dynamic, unpredictable conditions, as well as to further refine fail-safe mechanisms and fault tolerance strategies.

ACKNOWLEDGMENT

This paper is supported by the OPEVA project that has received funding within the Chips Joint Undertaking (Chips JU) from the European Union's Horizon Europe Programme and the National Authorities (France, Czechia, Italy, Portugal, Turkey, Switzerland), under grant agreement 101097267. The paper is also supported by Arrowhead PVN, proposal 101097257. Views and opinions expressed are however those of the authors only and do not reflect those of the European Union or Chips JU. Neither the European Union nor the granting authority can be held responsible for them. The work in this paper is also partially financed by National Funds through the Portuguese funding agency, FCT - Fundação para a Ciência e a Tecnologia, within project LA/P/0063/2020. DOI10.54499/LA/P/0063/2020, https://doi.org/10.54499/LA /P/0063/2020